\documentclass[a4paper, 12pt]{article}

\usepackage{graphicx}
\usepackage[export]{adjustbox}
\usepackage{amsmath} 
\usepackage{subfigure}
\usepackage{lscape}
\usepackage{authblk}
\usepackage{float}
\usepackage{array}
\usepackage{booktabs}
\usepackage{varwidth}
\usepackage{rotating}
\newcolumntype{C}[1]{>{\centering\let\newline\\\arraybackslash\hspace{0pt}}m{#1}}
\providecommand{\keywords}[1]{\textbf{\textit{keywords: }} #1}
\usepackage{verbatim}

\title{A data-driven model for Mass Media influence in electoral context}

\author[1]{Federico Albanese}
\author[2]{Claudio J. Tessone}
\author[3]{Viktoriya Semeshenko}
\author[4,5]{Pablo Balenzuela}

\affil[1]{Instituto de Investigaci\'on en Ciencias de la Computaci\'on, CONICET- Universidad de Buenos Aires, Argentina}
\affil[2]{URPP Social Networks and UZH Blockchain Center, Universit\"at Z\"urich, Andreasstrasse 15, CH-8050 Zurich, Switzerland}
\affil[3]{Universidad de Buenos Aires. Facultad de Ciencias Econ\'omicas. Buenos Aires, Argentina. CONICET-Universidad de Buenos Aires. Instituto Interdisciplinario de Econom\'ia Pol\'itica de Buenos Aires. Buenos Aires, Argentina}
\affil[4]{Departamento de F\'isica, Facultad de Ciencias Exactas y Naturales, Universidad de Buenos Aires, Av. Cantilo s/n, Pabell\'on 1, Ciudad Universitaria, 1428, Buenos Aires, Argentina.}
\affil[5]{Instituto de F\'isica de Buenos Aires (IFIBA), CONICET, Av. Cantilo s/n, Pabell\'on 1, Ciudad Universitaria, 1428, Buenos Aires, Argentina.}

\date{\today}

\begin{document}

\maketitle


\begin{abstract}
Mass Media outlets have historically occupied an important role in the political scenario and known to be persuasive in the process of opinion formation of citizens. 
The continuous availability of news related to the candidates and polls that precede an election day allow the monitoring and study of the relationship between Mass Media and behaviour of citizens. Based on this idea, we present a novel two-dimensional data driven model based on semantic analysis of newspapers and national election surveys, which we use to analyse how Mass Media as a single influence mechanism acts in order to give rise to the observed behaviour of the voters. Given a set of minimalist, yet sound assumptions, we were able to find a notable agreement between the model's predictions and the polls which help us to understand the underlying mechanisms of the interactions between readers and  media.

\end{abstract}
 
\section{Introduction}

\indent
\par Individuals get informed by consuming different formats of Mass Media (newspapers, radio, television, internet, etc).
The Mass Media play an important role in the formation of public opinion \cite{newman1990popular, ogden1913journalism, stivers2012media, yanich2004crime}. With its increased availability, this role constitutes a global phenomenon and has transformed the way individuals receive information on a daily basis. 

\par The access to the news and information can be understood within the agenda setting framework \cite{mccombs1972agenda}. 
Already one century ago, Lippmann observed that Mass Media dominates over the creation of pictures in our head, and that the public react not to the actual  events but to the pictures created in their heads \cite{lippmann1922}. 
The agenda setting process is in charge of remodelling the events that take place in our environment, and put them into an accessible model we interact with afterwards.

\par Agenda setting means the capability of Mass Media to bring issues concerned to the public and, concomitantly, to politicians. The basic claim is that as media devote more attention to an issue, the public perceives such issue as important. If the media bring up a specific topic, as for instance global warming, then they make the consumers think about it. This theory has been introduced  by McCombs and Shaw in their seminal study of the role of the media in the 1968 Presidential campaign in the US \cite{mccombs1972agenda}. 

\par Different approaches have been proposed to examine the role and influence of Mass Media on a society. On one side, social experiments have been designed, where group of subjects read news, and it was possible to directly measure how the opinions get modified afterwords.
In \cite{gerber2009does}, the authors shows that individuals who read the Mass Media constantly over time modify their political ideology and eventually vote explicitly in a different way. In \cite{king2017news}, it is demonstrated that exposure to news media causes U.S. citizens to take public stands on specific issues, join national policy conversations, and express themselves publicly more often than they would otherwise do.

\par On another stream of research, computational models have been implemented in order to  describe collective behaviour arising from interactions between Mass Media and the citizens \cite{pinto2016setting,gonzalez2006local, shibanai2001effects}, providing theoretical tools to understand micro to macro effects in this situations. Other approaches are based on the analysis of  large amount of data coming from surveys, as in  \cite{wanta2004agenda,oberholzer2009media}, from big data \cite{xie2018big} or from online traffic  \cite{yasseri2016wikipedia}, where the authors have shown that the relative change in the number of page views of a general Wikipedia page on the election day can offer a reasonable estimation of the relative change in turnout for that election at the country level.

\par In order to understand the changes produced by Mass Media in society, it is important to have a good representation of the ideology of the citizens. In \cite{brady2011art}, the author stress the importance of spatial diagrams of politics, because many fundamental problems of political science can be connected with them, and many different concepts (such as ideological constraint, cross-pressures, framing, agenda-setting, political competition, voting systems, and party systems, to name just a few) can be understood through spatial diagrams. In \cite{krasa2014policy}, for instance, Krasa and Polborn use a two-dimensional portrait with social and economics axes to represent a sample of the population in order to analyse political polarisation during elections in U.S.

\par If citizens can be placed in a two-dimensional diagram representing their ideology through answering a questionnaire as in \cite{PoliticalQuiz}  or \cite{compass2012political}, can Mass Media outlets be placed in same space? Elejalde et al. \cite{elejalde2018nature} analyse tweets to automatically compute the political and socio-economic orientation of news articles in order to represent the real and perceived bias of several Chilean media outlets.

\par In the present paper we go a step beyond, and combine some of the mentioned approaches to understand the role and influence that Mass Media may have on the opinions of citizens during political elections. We build a data-driven agent based model in which citizens are placed in a two-dimensional space according to their ideological positions as was done in \cite{krasa2014policy}. This framework allows to study different mechanisms in agent based models: social peer influence, mass media influence, etc. 
In particular, we focus on the hypothetical effects that a single isolated mechanism may have on a given population: the influence that Mass Media would have had on the intention to vote a given candidate.  In this context, we assume that the citizens only interact with news related to candidates in media outlets. Given the media outlets are not able to fill the ideological  questionnaire to be used afterwards to place them in the two-dimensional space like Nolan Charts for political spectrum diagram \cite{eysenck1954, Bryson1968} or Political Compass \cite{compass2012political}, we develop a novel method based on sentiment analysis  in order to represent the way different  media reflect the ideological positions of the candidates.

\par In order to test our approach, we compare the output of the model  with the results of $263$ national surveys conducted by different agencies obtained from Real Clear Politics \cite{RCP} obtaining a very good matching for a set of optimal parameters.

\par The paper is organised in the following way. In section \ref{sec:MaterialMethods} we present the data used and the data used in the study and the text mining tools applied to the article news. In section \ref{sec:scenario} we present the data-driven model used to represent the interactions between citizen and candidates as presented by mass media and finally, in section \ref{sec:results} we compare the output of the model with empirical data from polls during the political campaign. Finally in section \ref{sec:discussion} we discuss the utility of this modelling approach as well as future research lines.

\section{Data and Methods}
\label{sec:MaterialMethods}

\par In this section we describe the data sources employed in our analysis as well as the text mining techniques which were used to extract useful and relevant information from the news articles.  We retrieve data from three different mass media outlets, The New York Times, Fox News and Breitbart. We also use data from opinion polls in order to compare the output of the model.

\subsection{Mass Media Data}

\par We have selected three newspapers for the present analysis: New York Times (NYT), Breitbart and Fox News. The New York Times is the most searched online newspaper in all the states during the 2016 election campaign \cite{googleTrendsDiarios}, and has been classified as a democratic media \cite{berkeley2013MediaMap}. The Fox News portal is considered a Media Outlet with a republican bias, and the Breitbart portal was constantly quoted by Donald Trump and the media has made explicit his support since the beginning of the election campaign \cite{BrPorTrump}.

\par Thus, all the articles from the NYT, Fox News and Breitbart, corresponding to the electoral period from 07/28/2016 to 11/8/2016, were analysed (from the day of the last party convention of 2016, where formally candidates are defined, until the day of the election) containing at least the name of one of the two candidates: Hillary Clinton and Donald Trump.

\subsection{Polls}
\par In order to compare with the output of our model, time series from Real Clear Politics website \cite{RCP} were used. The data came from a total of 263 national surveys conducted by different agencies (an average of 2.7 surveys per day), in which the forecaster of vote of each candidate was measured in a gap of a few days (around 3-5 days). All national surveys that are presented in this work used a demographic balanced sample. 

The datasets supporting this article have been uploaded as part of the supplementary material.

\subsection{Sentiment Analysis}
\label{subsec:sentimentanalysis}

\par The sentiment analysis was performed utilising deep recursive models for the semantic composition applied to sentiment trees \cite{socher2013recursive}, in particular by the Stanford CoreNLP implementation of natural language processing \cite{manning2014stanford}. The algorithm consists of assembling a tree from the grammatical structure and a syntactic analysis of each phrase. Then, each word (node) is assigned a sentiment value, taken from a database: very positive, positive, neutral, negative or very negative. 
In addition, this algorithm takes into account if the words are intensifiers, appeasers, deniers, etc. Using deep machine learning techniques, the algorithm assigns a sentiment value  to each node starting from the inner nodes.  After several iterations, it ends up assigning the corresponding sentiment value to the root node which also corresponds to the total phrase. 
\par Even though there exists other alternative algorithms to perform sentiment analysis, such as those  based on the extraction of characteristics of the sentences \cite{doddi2014sentiment} or lexicon-based approaches to opinion mining \cite{taboada2011lexicon, muhammad2016contextual}, the Stanford CoreNLP is suitable for analysing our corpus of news given that each sentence is well-formulated with correct grammar and spelling and it uses this exact information to determine the sentiment of a phrase.

\section{The model}
\label{sec:scenario} 

\par A computational model is a simplified mathematical description of a real system which, by means of extensive computer simulations, describes natural and/or social phenomena \cite{melnik2015mathematical}. Different models have been built to help to understand the dynamics that governs the opinion formation process of a population \cite{Castellanoetal2009}, and to what extent a Mass Media has a direct affect on it. When the discussion is about a specific topic, this process can be modelled using one-dimensional approach based on classical models, as for instance the Deffuant’s model with continuous opinions \cite{pineda2015mass}, various extensions of the voter model \cite{ligget1995, masuda2015opinion}, and the Sznadj’s model \cite{Zhao2015,Crokidakis2012}.
On the other hand, when the intention is to model the whole agenda of a media outlet, typically a multidimensional approach is used based on Axelrod's model \cite{pinto2016setting, gonzalez2006local, axelrod1997dissemination, gonzalez2005nonequilibrium, rodriguez2009induced, rodriguez2010effects,mazzitello2007effects}. If it is intended to model opinion changes related to political positions, a bi-dimensional representation based in Nolan Charts political spectrum diagram \cite{eysenck1954, Bryson1968} or Political Compass \cite{compass2012political} could be used. This modelling approach, while sound, has not been performed until now. This step is a first contribution of our paper.

\par We developed a new framework to model the role and influence that  Mass Media can have and the dynamics of public opinion formation during elections as a single and isolated mechanism following a data-driven approach. 

\par We consider a population of $N$ citizens and $M$ Mass Media agents. Each citizen is represented by a coordinate in a bounded two-dimensional space which represents agents' characteristics and ideological preferences. In contrast with the work of Elejalde et al. \cite{elejalde2018nature}, here each Mass Media agent is described by two points, one for each candidate $c$, which represents the perception that the corresponding media outlet adopts about the candidate position. 
Different Mass Media outlet may depict differently the same candidates, and attribute them different positions depending on the news it decides to prevail, the emphasis to give to the topics and the sentiment of the article, as was shown in \cite{Albanese2019}. The idea is that a citizen, when he gets the news and information provided by Mass Media, does not interact with the a real candidate, but with the image reflected by each media outlet. That's why  we implement a sentiment analysis of news related with candidates, as we explain below.

\subsection{Position of the voters}

\par In order to place the citizens in the two-dimensional space we used the results obtained by Krasa and Polborn \cite{krasa2014policy},  where the authors take seven social and three economic questions from the National Election Survey of the year 2004 \cite{ANES} applied to $N=1066$ US citizens selected in a representative way. The social questions\footnote{The question number in the ANES survey corresponding to each one of the questions are: VCF0837, VCF0838, VCF0834, VCF0206, VCF0830, VCF0213 and VCF0130.} cover different topics: abortion, the role of women, discrimination against people of colour, the role of the state in helping ethnic minorities, the army and religion. The economic ones embrace the role of the state in the economy (state interventionism), unions, large companies and the average family income\footnote{The question number in the ANES survey corresponding to each one is: VCF0809, VCF0210 and VCF0209.}.  They assign a numerical range between 0 and 100 to the answers for each of the items, where the minimum stands for being disagreed with the subject and  the maximum to be totally agreed. 
The  obtained values are normalised between 0 and 1 and centred,  as shown by the green points in Figure \ref{Figura1}.

\par As far as we know, this representation, commonly used in sociology and human science \cite{brady2011art, mitchell2007eight, bell2012constitution, listhaug1990comparative}, has never been used before in agent based modelling.

\subsection{Position of the media}

\par Given that we assume the citizens only interact with news related to candidates in media outlets and  they are not able to fill the ideological  questionnaire to be used afterwards to place them in the two-dimensional space like Nolan Charts political spectrum diagram \cite{eysenck1954, Bryson1968}, or Political Compass \cite{compass2012political}, we develop a novel method based on sentiment analysis  in order to represent the way different  media reflects the ideological positions of the candidates.
This is a key ingredient of this model. Although media outlets can have a given ideological orientation \cite{elejalde2018nature}, our working hypothesis is that citizens get informed about the candidates through Mass Media. That is why we assume the relevant ingredient to capture the essence of the interaction between the media and citizens in a context of elections is the image of the candidates that media project. The way we find to do this is through a sentiment analysis of the media outlets related to the key concepts that allow choosing the coordinates in a two-dimensional diagram with social and economic axes.

\par We use a dictionary of keywords based on social and economic questions and  recursive deep models for semantic composition over sentiment treebank  in order to detect the sentiment of a given sentence \cite{socher2013recursive, manning2014stanford}, and quantify positively or negatively the context where these keywords appear. 

\par
\begin{figure}[H]
\centering
\includegraphics[width=1.0\linewidth]{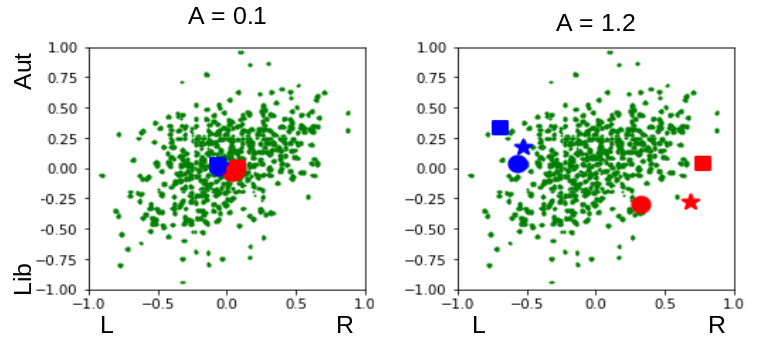}
\caption{{\bf Citizen and media representation in the two-dimensional social-economic space}: The initial positions of the citizens and Mass Media outlets for different numerical values of scaling factor $A$. Plots : citizens (green), 3 Mass Media outlets (Trump in red and Clinton blue). The big circles, the stars and squares correspond to the position for the New York Times, Fox News and Breitbart perception of the candidate respectively. Lib and Aut represent libertarian and authoritarian in the vertical social axis; and L and R represent left and right in the horizontal economic axis.}
\label{Figura1}
\end{figure}

 The following steps have been done for locating the Mass Media agents in the two-dimensional space:

\begin{description}

\item [Create dictionary.] We define four dictionaries\footnote{The list of words of the four dictionaries can be found in the Supplementary material.} containing words for the libertarian, authoritarian, left and right topics (each of the semi-axes of the plane), using the words extracted from questions in the ANES polls, Political Compass \cite{compass2012political} and IsideWith \cite{isidewith}. These last two surveys were added in order to complete dictionaries with a larger number of words, and get a better statistics in the semantic analysis.

\item [Learning.] In order to know if a phrase contributes to a given semi-axis of the bi-dimensional representation, we identify keywords of one of the four semi-axes and try to know if they are mentioned positively or negatively. We do that using recursive deep models for semantic composition over sentiment treebank applied to this phrase

\item [Sentiment Analysis.] Each sentence $i$ gets an assigned value $l_j (i)\in[-2,2]$, for each Media and for each candidate, where $j$ corresponds to one of the four topics or semi-axes: $j=r,l,aut,lib$ (right, left, authoritarian and libertarian respectively). Neutral phrases will have $l_j=0$ since there are not in favour nor against the statement represented by the list of words of a given semi-axis. Very negative and very positive ones will have $l_j=-2$ and $l_j=+2$ respectively, and will be twice of the weight of negative (-1) or positive (+1) sentences. Given this notation, for instance, $l_{left}(i)$ is the sentiment value for the sentence $i$ of the economic left list.

\item [Calculate coordinates.] The information collected in these lists is used to calculate the coordinates in the two-dimensional space representing the candidate $c$ ($c=C$ for Clinton and $c=T$ for Trump) from the perspective of a given Media Outlet $m$ (the $x_{m,c}$ coordinates for economic and $y_{m,c}$ for the personal axis):

\begin{equation}
x_{m,c} = \frac{\sum_{i=1}^{n_r} l_{r}(i) - \sum_{i=1}^{n_l} l_{l}(i)}{\sum_{i=1}^{n_r} \mid l_{r}(i) \mid + \sum_{i=1}^{n_l} \mid l_{l}(i) \mid }
\label{eqMODELO1}
\end{equation}

\begin{equation}
y_{m,c} =  \frac{\sum_{i=1}^{n_{aut}} l_{aut}(i) - \sum_{i=1}^{n_{lib}} l_{lib}(i)}{\sum_{i=1}^{n_{aut}} \mid l_{aut}(i) \mid + \sum_{i=1}^{n_{lib}} \mid l_{lib}(i) \mid}
\label{eqMODELO2}
\end{equation}

\par where  $n_{r}$, $n_{l}$, $n_{aut}$ and $n_{lib}$ are the number of phrases in right, left, authoritarian and libertarian categories. The values $x_{m,c}$ and $y_{m,c}$ are normalised and centred using the same method that was used before for the position of the voters.

\item [Distance. ] Both, the citizen and the Mass Media are placed in the same two-dimensional space. However, they have different scales due to the fact that their positions were obtained with different methods. This is the reason why a parameter of scale $k$ was introduced to our model in order to take into account this issue. Therefore, the coordinates $x_{m,c}$ and $y_{m,c}$ are multiplied by a constant $k$ and $A_m=k \sqrt{(x_C-x_T)^2+(y_C-y_T)^2}$ is defined as the distance between the centre of mass of the location of one candidate and another. The final result for different values of $A=<A_m>$ is plotted in Figure \ref{Figura1}.

\end{description}

\par Figure \ref{Figura1} also shows the representations of the two candidates for each of the three Mass Media outlets. It is interesting to emphasise that in this representation the candidates are perfectly grouped by media and by candidate. It is observed that Trump’s representations for each of the Media are in the lower right corner, whereas all of Clinton’s points are in the upper left side. This representation of the Democratic candidate as authoritarian and more of economically to the “left” than Donald Trump is consistent with the article's highlight of her tax reform, where the wealthiest should contribute to a greater extent. On the contrary, Donald Trump opposed such reform and also the national health plans for all citizens, which is informally named ``Obamacare''.

\par On the other hand, Figure \ref{Figura1} also shows how the candidates are aligned according  to  each  media.   In both cases Breitbart provides the most authoritative representation, followed then by Fox News and the New York Times. The  fact  that  the  points  are  grouped  both  by  candidates  and  the media indicates that proposed method was able to distinguish effectively  the differences  between  the  candidates  using  the  information  available  in the news only.

\par Curiously, it could be seen in Figure \ref{Figura1} that the Mass Media agents align themselves in a perpendicular axis with respect to the rest of the individuals. In order to understand this phenomenon,  a game theoretic approach could be invoked. For a two-dimensional space where two entities compete for a good distributed uniformly inside a square in the plane and where agent keeps the goods which are closer to himself than to the other agent, the Nash Equilibrium is found when entities have the same position at the centre of the square \cite{easley2010networks}. However, the two candidates can not be located in exactly the same place due to the characteristics of the scenario, where the candidates tend to polarise \cite{stonecash2003diverging, layman2001great}. Here, the polarisation is represented in a two-dimensional opinion space as two agents in different, antagonistic positions. Therefore, they theoretically must polarise in a perpendicular direction to the axis where the population is mostly distributed in order to maximise their votes but also keep a distance with the other candidate. Exactly this phenomenon emerges naturally from the text analysis proposed, and it is observed in Figure \ref{Figura1}, validating the methodology.

\par Finally, we validate both methodologies in order to be sure we obtain consistent results. The details of these procedures will be clarified in the next section.

\subsection{Sentiment Analysis and ANES: validation}

\par As previously mentioned, locating Mass Media and citizens in the same two-dimensional space has an intrinsic difficulty: the same methodology can not be used for both. The reason behind this statement is that the position of the citizens can be defined using the analysis of a survey, whereas it cannot be used for Mass Media. In order to validate the methodology used, we compared the  positions in the two-dimensional space assigned on all possible answers of the chosen survey ANES questions  (used for citizens) with the positions based on the sentiment analysis methodology (used for Mass Media) for the same answers.

\par
\begin{figure}[H]
\centering
\includegraphics[width=1.0\linewidth]{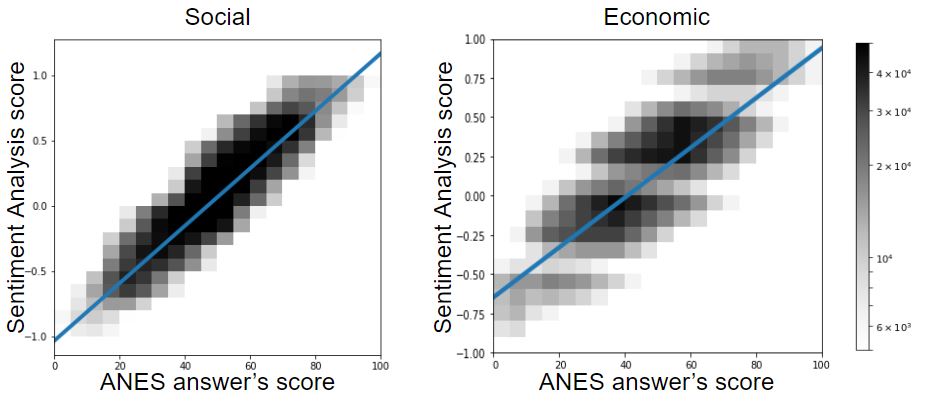}
\caption{{\bf Validation of representation methods}: Comparison of all possible combinations of answers to the social (left) and the economic (right) questions of ANES between both methodologies: scores and sentiment analysis. In blue there are lineal fits with slopes equal to $0.022$ and $0.015$ respectively and p-values lower than $0.001$.}
\label{Figura2}
\end{figure}

 Texts were prepared with multiple choice answers to each of the ANES questions. Then, we applied the same procedure to the texts that are used to position the media based on the sentiment analysis of news articles. If both methodologies were equivalent, the value assigned by each of the methods would have resulted the same.  In Figure \ref{Figura2} we plot the combination of all possible answers to social and economic questions. It can be seen that the linear relation validates the correspondence between the methods.

\subsection{Dynamical Rules}

\par Given that one of the goals of this work is to study the effects that the most important mass media would have on a measurable social behaviour (as the polls in the electoral period previous to elections), we assume that citizen agents in our models only interact with the news related to the candidates, represented with the media agents as explained above. The rationale of this approach is to study the effects of a single isolate mechanism.

\par
\begin{figure}[H]
\centering
\includegraphics[width=1.0\linewidth]{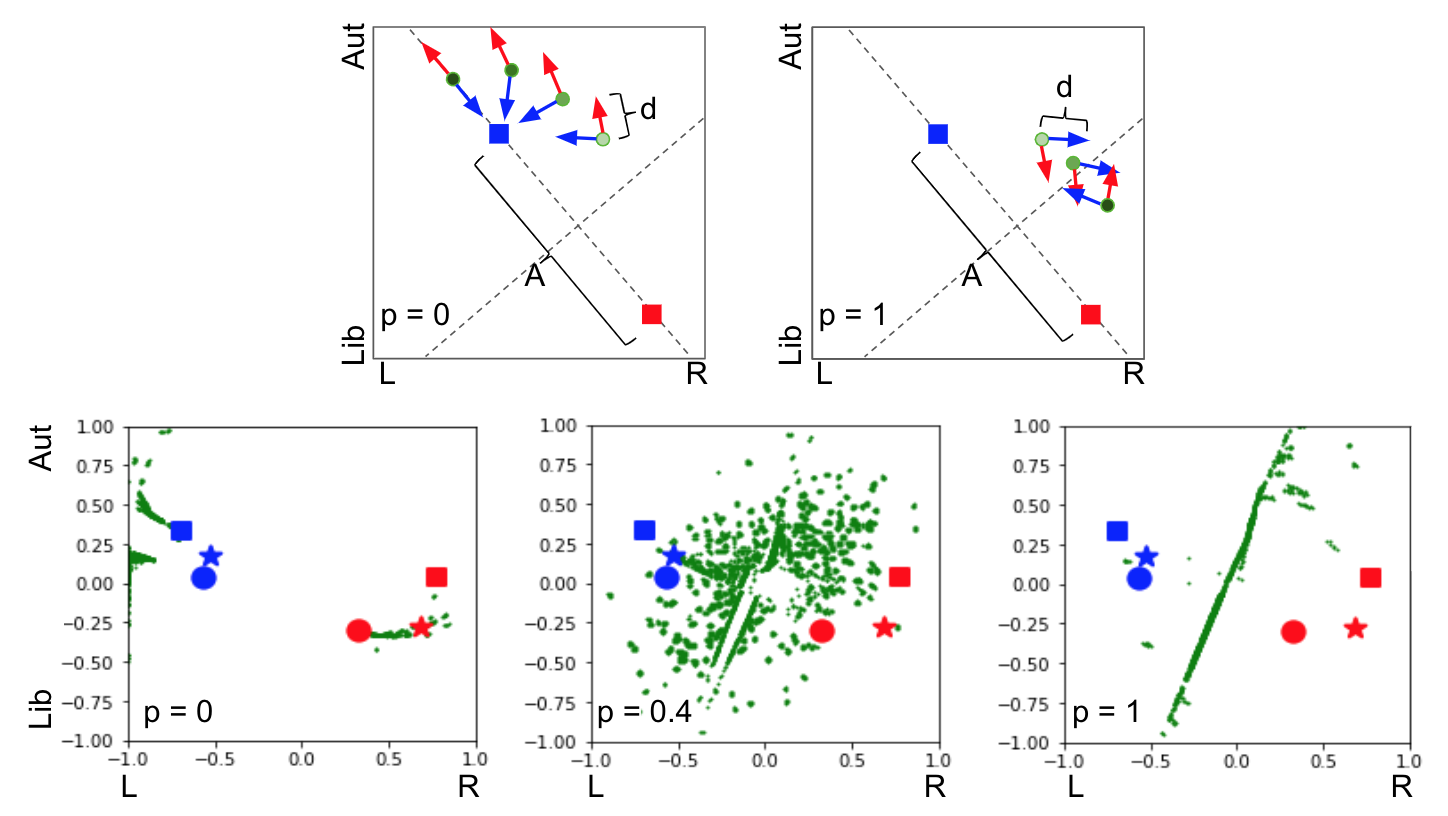}
\caption{{\bf Dynamical rules}: Representation of two candidates (squares Blue and Red) and a citizen. The arrows indicate the possible movements of the agent when interacts with the media representation of a candidate. Lib and Aut represent libertarian and authoritarian in the vertical social axis; and L and R represent left and right in the horizontal economic axis. On the top the possible movements of an agent with an attractive dynamic ($P=0$) and on the bottom, a repulsive dynamic ($P=1$). The shades of green represent how an agent would move through time in the two-dimensional space.}
\label{Figura3}
\end{figure}

 We make the following set of assumptions in order to establish rules to model the process of Mass Media social influence:

\begin{itemize}
    \item Each citizen interact only with one Mass Media outlet in a period. Thus, one interaction is the representation of an individual consuming a media outlet's content one single day.
    \item A citizen interacts with the Mass Media which results to be closer to his/her preferences (it corresponds to the closest distance to the line that connects both media points). 
    \item A citizen reads news related to both candidates in each period. It means that the agent interacts with both points corresponding to a given Mass Media outlet.
    \item A citizen  reacts differently depending on whether he/she interacts with his/her preferred candidate or the opposite one:
        \begin{enumerate}
            \item If the agent interacts with the preferred candidate (the closest one), with probability $(1-p)$ he will be attracted to the candidate by a distance $d$, and with probability $p$ repelled  (blue arrows in top panels of  Figure \ref{Figura3}).
            \item If the agent interacts with the opposite candidate (the furthest one), with probability $(1-p)$ he will be
             repelled  by a distance $d$, and with probability $p$ attracted (red arrows in  top panels of  Figure \ref{Figura3}).
         \end{enumerate}
         \item The distance $d$ is given by the following equation:
            \begin{equation}
                d = d_0 m_c(t) x,
            \end{equation}
            where $d_0$ is a model parameter and quantifies the degree of influence exerted by the Mass Media outlet on the reader, $x$ is a random variable which takes values $x=-1$ with probability $p$ and $x=+1$ with probability $(1-p)$, and $m_c(t)$ is a data-driven parameter which counts the number of phrases that contain the candidate $c$ in the news in the given Mass Media outlet in the period $t$. Consequently, $m_c(t)$ is a measure of how much each candidate is mentioned. The larger $m_c(t)$ the more intense the interaction will be. 
            
        \item After interacting with both candidates, the citizen gets closer (positive influence) to his/her preferred candidate with probability $(1-p)$, and moves away from him (negative influence) with probability $p$. Note that ``positive influence" in this model is composed by attraction to his/her preferred candidate and repulsion from the opposite candidate.
        
\end{itemize}

In order to get insights about the behaviour produced by the sketched  dynamical rules, lets focus on some specific cases: 

\begin{itemize}
    \item If $p=0$, individuals get attracted to their preferred politician and repel from the opposite one, emulating an attractive dynamics towards the candidates, as sketched for a single agent in top left panel of  Figure \ref{Figura3}. This dynamics polarises the population, as shown in left lower panel of  Figure \ref{Figura3}. Once an agent has approached a candidate, it makes it difficult to get away when positive interactions predominate.
    \item  If $p=1$, a repulsive dynamic is present and the citizens move away from the preferred candidates and approach the one that are distant. Logically, majority will end up in the middle between the two candidates, as shown for a single candidate in top right panel of Figure \ref{Figura3} and at population level at low right panel of the same figure.
\end{itemize}

\section{Output of the model}
\label {sec:results}

\par In this section we run the model in order in order to analyse the dynamics of the model. 

\par The dynamics of the model depends on the parameters $A$, $d_{0}$, $p$ and $\tau$. 
The parameter $A$ gives the scale relation between metrics used to locate the citizen and the one corresponding to the Mass Media agents in the plane (see Figure \ref{Figura1}).
The parameter $d_0$ is a measure of influence of a Mass Media on a citizen (if $d_0$ is high, the citizen performs a larger change in its ideological position after reading the Mass Media), and $p$ is the probability that an agent moves away from his voter's preference (negative influence). The parameter $\tau$ takes into account the possibility that the degree of influence between Mass Media and citizens could not be instantaneous but mediated by a lag $\tau$.

\par In order to infer the percentage of citizens electing each candidate from our model, we
assume that citizens vote (or explicitly manifest their support) to the closer one in the ideological space. However, it should be taken into account the fact that the US adult citizens are not enforced to vote, so there is a region of undecided voters corresponding to $ 44 \% $ of the population (the participation rate in the 2016 presidential election \cite{porcentajePArticipacion}. Then we simply take the  $ 56 \% $ of the closest citizens to each candidate and construct the time series of agents supporting for Clinton or Trump.

\par We proceed to determine the best performance of the model by comparing the time series generated by the model and the results of the polls. The comparison consists in minimizing the  absolute average distance between the curves (difference Clinton-Trump) and maximizing the correlation in the four-dimensional parameter space given by $p$, $ d_{0} $, $A$ and $\tau$.

The range of variations of each parameter is the following:
\begin{itemize}
    \item $d_0$ varies between $0.001$ and $0.1$.  For values greater than $0.1$ the agents move at larger distances in a single interaction producing large oscillations that are not observed in data. On the other hand, values lower than $0.001$ produce negligible displacements which are not interesting for our analysis.
    \item The probability parameter $p \in [0,1]$.
    \item $A$ is varied between $0.6$ and $1.5$. The smaller values set the media agents too close to each others and are not interesting to be considered, while bigger values set them outside the boundary box.
    \item The lag $\tau$ takes values between 0 and 20 days.
\end{itemize}

 After a  complete grid search is performed, we look for a combination  of  parameters  that  maximise the correlation and also minimise the distances between the mentioned series. The optimal performance corresponds to the set: $ d_{0} = 0.04$; $p = 0.2$; $\tau = 10$; $A = 1.5$. 
 
\par The optimal set of parameters found is worthy to be interpreted as we consider they highlight the  importance of the model. The small value of $p$ ($p=0.2$) indicates that the interactions between the Mass Media and the readers are mainly positive (either by supporting the preferred candidate or repelling from the opposite). This result  is consistent with previously published research \cite{gerber2009does,yasseri2016wikipedia}. On the other hand, the computational model gives its best estimations $10$ days in advance since a maximum performance could be achieved with $\tau = 10$ days, in line with delayed correlations between news influence and polls found in \cite{Albanese2019}.  As for the parameters $A$ and $d_{0}$, it is interesting that the best model corresponds to the largest possible value of $A$ ($A=1.5$) and a relative small $d_{0}$ ($d_{0} = 0.04$). In reality, candidates tend to separate themselves from the opponents because polarisation is a common strategy in a two-party system \cite{layman2006party, stonecash2003diverging,layman2001great}. 

\par
\begin{figure}[H]
\centering
\includegraphics[width=1.0\linewidth]{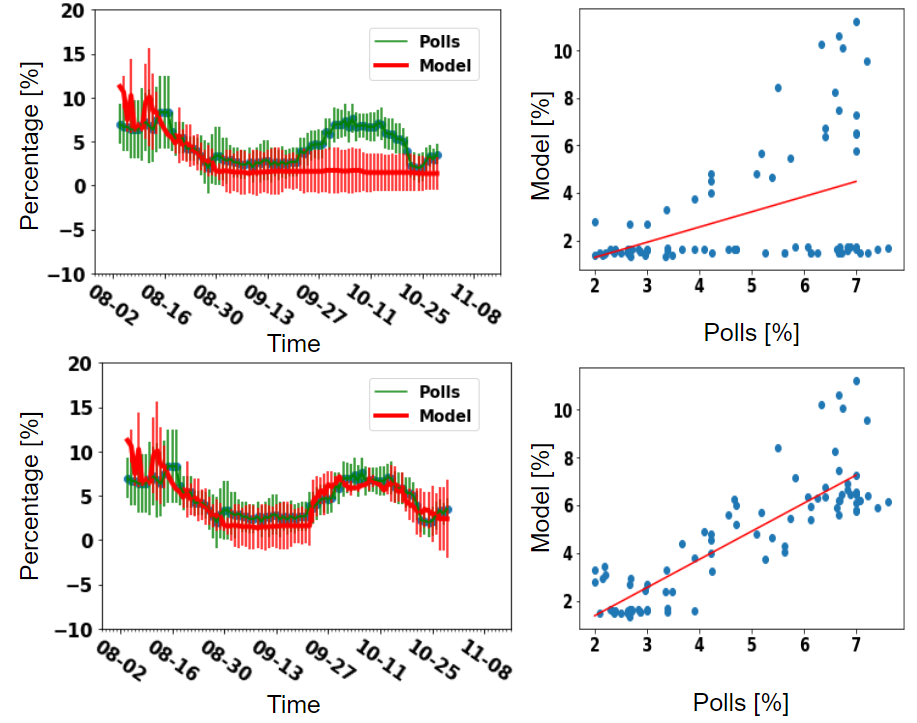}
\caption{{\bf The time series produced by the model and the polls}: The time series with the differences in percentage of voters (Clinton - Trump) produced by the model and of the polls through time (left) and the pair scatter plot (right). The top panels correspond to the model with $ d_{0} = 0.04$; $p = 0.2$; $\tau = 10$; $A = 1.5$. The bottom panels correspond to $ d_{0} = 0.04$; $p = 0.2$; $\tau = 10$; $A = 1.5$ for the first part and $p=0.4$ and $d_{0}=0.01$ for the last month.}
\label{Figura4}
\end{figure}

 In top left panel of  Figure \ref{Figura4}, the superposition of the time series of the polls and the model ($ d_{0} = 0.04$; $p = 0.2$; $\tau = 10$; $A = 1.5$) is observed. Since the model has a random variable, multiple iterations where performed and the error bars were assigned. This computational predictions based on the text of news articles are consistent with the polls for the first half of the electoral period. The top right panel of Figure \ref{Figura4} shows the difference in the percentage of voters (Clinton - Trump) produced by the model vs those produced by the polls. The points grouped in the horizontal line correspond to the region in which the model does not fit the data.

\par However, this implementation does not replicate correctly the shape of the poll's curves in the period of four weeks before the election day. This result could be related to the increasing of  ``negative advertisement'' given that it could be used as a last resource in order to shorten the distance between the candidates \cite{peterson2005primary}. Also, individuals who are interested in politics might have already made his/her election (and therefore are not willing to change it), and those who are not politicised may be influenced less by news due to a lack of care. Consequently, owing to this change in the behaviour, different parameters could be needed in order to fit the last month. If we increase the value of the parameter $p$ (representing  more negative interactions) and decrease the value of $d_{0}$ (representing weaker social influence) we get a new optimal fitting for the last month with $p=0.4$ and $d_{0}=0.01$. We show the full curves (with $ d_{0} = 0.04$; $p = 0.2$; $\tau = 10$; $A = 1.5$ for the first part and $p=0.4$ and $d_{0}=0.01$ for the last month) in low left panel of Figure \ref{Figura4}.

\par Also, it is easier to observe in top-right panel of  Figure \ref{Figura4}  that the model is producing an accurate fit of polls in the first $10$ weeks but fails in the last $4$ weeks. The improvement of allowing a change in behaviour in the last weeks could also be seen in right panels of Figures \ref{Figura4} (top and down), where slope is closer to 1 and the fit is statistically better in the second case.

\section{Discussion}
\label {sec:discussion}

\par In this paper, we proposed a novel framework to bound the influence Mass Media can have on individuals' opinion formation process. We build a data-driven model to study the hypothetical effects that a single isolated mechanism would have on a given population (in this case, Mass Media influence). 
This model is based on a representative sample of a population (citizens agents) placed in a bounded socio-economical two dimensional space \cite{krasa2014policy}  and a representation of the candidates as portrayed by Mass Media in the same space. 

The novelty of our approach lies in the following hypothesis: the citizens get informed about the candidates through  Mass Media  and react accordingly. That is why we assume the relevant ingredient to capture the essence of the interactions between the media and citizens in a context of elections is the image of the candidates as projected by media. The way we find to do this is through sentiment analysis of media outlets related to the key concepts that allow choosing the coordinates in the mentioned two-dimensional diagram with social and economic axes.
 
 \par The dynamical rules are chosen in a simple way, assuming that the citizen reads news related to both candidates and gets attracted to his/her preferred candidate or repelled from the opposite one with probability $(1-p)$, which can be considered as a positive influence towards its own candidate. 

\par The proposed model and its optimal parameters are consistent with the literature for negative propaganda \cite{peterson2005primary}, polarisation strategies \cite{layman2006party, stonecash2003diverging, layman2001great} and the influence of the Mass Media \cite{gerber2009does, oberholzer2009media}, replicating those behaviours.

\par It should be noted that the model assumes citizens approach or move away from the image that the Mass Media reflects about the candidates when expressing their public opinion, but such changes should not necessarily be permanent or may reflect a difference between the manifestation of their vote and the true ideological position. That is, the observed changes in the model related to the positions of the citizens should not be seen as permanent changes in their ideological position, but as transitory commitments with the two majority voting options.

\par Finally, we would like to remark that this kind of framework could be taken as a starting point for data-driven modelling of other mechanism of  social influence in socio-economical environments. 





\section{Competing interests}

The authors declare no competing interests.

\section{Authors' contributions}

F.A. collect the data, made the calculations and participate in the writing of the manuscript. C.T, V.S. and P.B. design the study, discussed and interpreted the results and participate in the writing of the manuscript.

\section{Acknowledgements}

We acknowledge Sebastian Pinto for carefully reading and interesting discussions.



\section{Figures and table captions}

\newpage

\newpage

\appendix

\section{Supplementary material: Dictionaries for the position of the Mass Media}


The four dictionaries were defined from the questions of the ANES poll, Political Compass and IsideWith and were used in order to find the position of the Mass Media agents. By no means, this work suggests that the listed words belong intrinsically to their semi-axis. On the contrary, the words were chosen from the questions depending on whether they appeared in a context where the affirmative answer was in favor of the semi-axis. This is the reason why there are terms such as 'employees' which appears both in Right and Left dictionaries. In this example, 'employees' clearly defines an economic axis, but can be used for eider of them.

{\bf Right}:
[market share,market, Bank, Banks, China, Free Trade Agreement, NAFTA, TPP, Trans-Pacific Partnership, bonds, businesses, charity, company, controlling inflation, corporation, corporations, corporations, corporation, currency, disadvantaged, earned, employees, employers, free market, freedom, full-time employees, highly taxed, imported products, increase the tax, individual freedom, industry, inflation, invest, job, jobs, legal currency, liberty, manipulate money, market, money, nationality, offshore bank accounts, offshore, increased, reduce, open market, paid, pay, personal fortunes, private companies, privately, privately managed accounts, profit, profits, property, property taxes, raise taxes, real estate, recession, required, rich, salaried employees, sales tax, same job, sellers, shareholders, stocks, successful corporations, tax, tax incentives, tax rate, taxes, taxpayers, the economy,economy, the rich, the same salary, salary, same salary, trans-national corporations, unemployment]

{\bf Left}:
[Bank, Banks, Federal Reserve Bank, basic income, big government, businesses, class, corporations, economic globalisation, economic stimulus, employees, employers, environment, for the people, full-time employees, government, income, income program, increase the tax, industry, job, jobs, labor unions, medical care, minimum wage, more restrictions, obamacare, paid, pay, penalise businesses, pension payments, pension plans, profits, property taxes, protectionism, public funding, public spending, public, increased, reduce, raise, recession, reduce debt, regulation, require regulation, required, restrictions, salaried employees, sales tax, same job, serve humanity, social insurance, social plan, social security, subsidise, subsidise farmers, tax, tax business, tax rate, taxes, the economy, economy, the government, the national debt, national debt, the same salary, salary, same salary, unions, universal basic income, wage, workers]

{\bf Libertarian}:
[allow, Planned Parenthood, abortion, adoption, adoption rights, anarchism, anti-discrimination, anti-discrimination laws, artist, assisted suicide, be allowed, black lives matter, child adoption, civil liberties, classroom attendance, contamination, cultures, democracy, democratic, democratic political system, different cultures, discriminate, discrimination, drugs, environment, free, free birth control, freedom, gay couples,gay, gender identity, health insurance,insurance, homosexual, homosexuality, immigrants, immigration, keep secretes, legal, legalization, legalise, liberty, marijuana, naturally homosexual, openness about sex, personal use, poet, pollution, pornography, possessing drugs, possessing marijuana, privacy, private, pro choice, rehabilitation, same sex, same sex couple, same sex marriage, same sex relationship, secretes, societys support, transgender, transgender people]

{\bf Authoritarian}:
[combat roles, the U.S. Military, AntiFa, Confederate, Confederate flag, Confederate monuments,Confederate memorials, First-generation immigrants, God, Multinational companies, accept discipline, allowed to reproduce, army, authority, businessperson, capital punishment, catholicism, church, command, commanded, confederate, counter-terrorism, country, country of birth, crime, criminal, criminal justice, criminal offence, death penalty, death penalty, military, deny service, discipline, discriminate, discrimination, domestic terrorist organization, education, establishment, fully integrated, government control, homophobic, immigrant, immigration, judge, manufacturer, marital rape, maturity, military, military action, moral, nation, nationalism, nationality, non-marital rape, obey, obeyed, official surveillance, one-party state, population control, prison, prisoner, prisoners, pro life, punished, punishment, race, religion, religious, religious beliefs, religious values, soldier, superior race, surveillance, terrorism, terrorist, terrorist organization, war]

\section{Supplementary material: Analysing an article in order to define the position of the Mass Media Agents (an example)}

As it is mention in Section 2.2 (Position of the media), the position of the media can't be determined with the same methodology that was used for the agents. Consequently, a semantic analysis of the articles was needed in order to know how the Mass Media portrays each the ideology of a candidate.

The complete list of questions from which the terms of the dictionaries where extracted and the dictionaries itself that were used can be also found in the Supplementary material. As it is described in the main article before sentiment analysis algorithms are applied to the sentences of an article that contain at least one word from one of the dictionaries. In this section an example of this procedure, is shown.

The example article is: Trump Said Women Get Abortions Days Before Birth. Doctors Say They Don’t. (www.nytimes.com/2016/10/21/health/ donald-trump-debate-late-abortion-remarks.html)  from the New York times on  the $10/21/2016$ (and a equivalent procedure was implemented for all the articles of the three Mass Media). In this particular case, an example of the sentences that contain a term of a dictionaries and their corresponding output of the sentiment analysis are:

\begin{itemize}
    \item ``In the presidential debate Wednesday night, Donald J. Trump expounded on pregnancy and abortion , asserting that under current abortion law, you can take the baby and rip the baby out of the womb in the ninth month, on the final day. ''. A sentence that contain the term abortion from the libertarian dictionary. The sentiment for this sentence is negative (-1).
    \item ``A few wrote emotionally about their own late-term abortions , and said that Mr. Trump minimised the pain they felt in having to make one of the most difficult decisions in their lives''.  A sentence that also contain the term abortion from the libertarian dictionary. The sentiment for this sentence is negative (-1).
\end{itemize}

So, for this sentences the lists: $l_{right}$, $l_{left}$,$l_{authoritarian}$ and $l_{libertarian}$ will be respectably: [0,0],[0,0],[0,0] and [-1,-1] since both of them are negative and are statements about liberal ideas.

After this procedure, the steps continues as described in the Section 2.2 .

\end{document}